\documentclass[12pt,a4paper]{article}
\usepackage[utf8]{inputenc}
\usepackage{amsmath,amsfonts,amssymb,amsthm, multirow}
\usepackage{graphicx}
\usepackage{fullpage}
\usepackage{hyperref}
\usepackage{natbib}
\usepackage{color}
\usepackage{tabularx}
\usepackage{amsfonts,amssymb,amsmath,amscd,latexsym,dsfont}
\newcommand{\balph}{\boldsymbol{\alpha}}
\newcommand{\bom}{\boldsymbol{\omega}}
\newcommand{\lp}{\ell_{\text{pen}}}
\newcommand{\bOm}{\boldsymbol{\Omega}}
\newcommand{\bth}{\boldsymbol{\theta}}
\newcommand{\bmu}{\boldsymbol{\mu}}
\newcommand{\bSigma}{\boldsymbol{\Sigma}}
 \newcommand{\bm}{\boldsymbol{m}}
 \newcommand{\M}{\mathcal{M}}
 \newcommand{\mO}{\mathbf{O}}
\newcommand{\bx}{\boldsymbol{x}} \newcommand{\tx}{\textbf{x}}
\newcommand{\tz}{\textbf{z}} \newcommand{\bz}{\boldsymbol{z}}
 
\newcommand{\MICL}{\text{MICL}}
\newcommand{\pdf}{f}

\newcommand{\btau}{\boldsymbol{\tau}}
\newcommand{\tzstar}{\textbf{z}^{\star}}
\newcommand{\bTh}{\boldsymbol{\Theta}}
\newcommand{\mhat}{\hat{\boldsymbol{m}}}

\newcommand{\BIC}{\text{BIC}}
\newcommand{\argmax}{\mathop{\mathrm{arg\,max}}}

\title{Variable selection for mixed data clustering: a model-based approach}
\author{Matthieu Marbac and Mohammed Sedki}

\begin{document}
\maketitle

\begin{abstract}
We propose two approaches for selecting variables in latent class analysis (\emph{i.e.,} mixture model assuming within component independence), which is the  common model-based clustering method for mixed data. The first approach consists in optimizing the BIC with a modified version of the EM algorithm. This approach simultaneously performs both model selection and parameter inference. The second approach consists in maximizing the MICL, which considers the clustering task, with an algorithm of alternate optimization. This approach performs model selection  without requiring the maximum likelihood estimates for model comparison, then parameter inference is done for the unique selected model. Thus, the benefits of both approaches is to avoid the computation of the maximum likelihood estimates for each model comparison. Moreover, they also avoid the use of the standard algorithms for variable selection which are often  suboptimal  (\emph{e.g.} stepwise method) and computationally expensive. The case of data with missing values is also discussed. The interest of both proposed criteria is shown on simulated and real data.
\end{abstract}

\section{Introduction}
Clustering permits to summarize large datasets by grouping observations into few homogeneous classes. Finite mixture models \citep{McLachlan:04,McNicholas2016} allows assessment of this unknown partition among observations. They permit dealing with continuous \citep{Ban93,Cel95,morris2016clustering}, categorical \citep{Mei01,mcparland2013clustering,marbac2016}, integer \citep{karlis2007finite} or mixed data \citep{browne2012model,kosmidis2015mixture,marbac2015copules}. When observations are described by many variables, the within components independence  permits achievement of the clustering goal, by limiting the number of parameters \citep{Goo74,Han01,Mou05}. Like in regression \citep{Davis20113256,Huang20062020} or classification \citep{Greenshtein2009385,HUANG1994205}, variable selection should be done in clustering. Indeed, in many cases, the partition may be explained by only a subset of the observed variables \citep{biernacki2015}.  So, by performing a selection of the variables in clustering, both model fitting and result interpretation are facilitated. Indeed, for a fixed sample size, selecting the variables improves the accuracy of parameters and class identification. Moreover, such method brings out the variables characterizing the classes, thus facilitating the interpretation of the clustering results.

The first approaches for selecting the variables have been developed to cluster continuous data. Thus, \citet{Tadesse:05} consider two types of variables: 
the set of the \emph{relevant variables} (having a different distribution among components) and the set of the \emph{irrelevant variables}  (having the same distribution among components) 
which are independent of the relevant ones. This method has been extended
 by considering a set of redundant variables. The distribution of the redundant variables is modelled by linear regressions on the whole discriminative variables \citep{Raftery:06} or on a subset of the discriminative variables \citep{Maugis:09b}. Authors propose to perform model selection by maximizing the Bayesian Information Criterion (BIC, \citet{Schwarz:78}). However, this maximization is complex because many models are in competition, and because each model comparison requires calls of EM algorithm to obtain the maximum likelihood estimates (MLE). This optimization can be carried out by a greedy search algorithm. This algorithm converges toward a local optimum in the space of models, but there is no guarantees that this optimum is the global one. This algorithm is feasible for quite large datasets, but it is computationally expensive when many variables are observed. Considering the latent class model \citep{Goo74}, \citet{dean2010latent} then \citet{white2016bayesian} introduce a similar way for selecting variables in a categorical data clustering.

For the first contribution of this paper, we show how to select the variables, according to the BIC, for a model-based clustering of mixed data. The model  considers two kinds of variables (relevant and irrelevant) and assumes within component independence.  Note that this model is useful
especially when the number of variables is large \citep{Han01}, that is the most common situation where variable selection is needed. The within components independence allows us to easily implement a modified version of the Expected-Maximization (EM) algorithm proposed by \citet{green1990use}, which permits the maximization of the penalized likelihood. Thus, the proposed method permits the selection of variables in clustering of mixed data according to any likelihood-based information criterion, like the AIC \citet{Aka73} or the BIC. Other penalised criterion considering the complexity of the model space could also be optimized \citep{massart2007concentration,meynet2012selection,bontemps2013clustering}.

The BIC approximates the logarithm of the integrated likelihood by adding a term of $\mathcal{O}(1)$. This term can deteriorate its performances, when few observations are available. Moreover, it does not focus on the clustering goal, thus the Integrated Complete-data Likelihood criterion (ICL) has been introduced by \citet{Biernacki:00}. This criterion makes a trade off between the the model fit to the data and the component entropy. Moreover, when the components belong to the exponential family and when conjugate prior distributions are used, this criterion does not imply any approximation.  \citet{Biernacki:10} shows that this exact criterion can outperforms the BIC. However, selecting the variables according the ICL is complex. Therefore, \citet{marbac2015variable} introduced the Maximum  Integrated Complete-data Likelihood criterion (MICL) for selecting  variables of a diagonal Gaussian mixture model.  The ICL and the MICL are quite similar, because both of them are based on the integrated complete-data likelihood. However, the MICL  uses the partition maximizing this function, while the ICL uses the partition provided by a MAP rule associated to the MLE. 

For the second contribution of the paper, we show that the MICL keeps a closed form for a mixture model for mixed data, if prior distributions are conjugate. Hence, model selection is carried out by a simple and fast procedure which alternates between two maximizations, for providing the model maximizing the MICL. We shows that this exact criterion (\emph{i.e.,} not implying any approximation) can outperform the asymptotic criteria (like the BIC). Finally, we show that the two contributions of this paper improve the clustering results when data have missing values. To manage missing values, we assume that values are missing at random \citet{Little:2014}. 

Section~\ref{sec:model} presents the mixture model for mixed data.
Section~\ref{sec:BIC} details the selection of variables according to the BIC, while Section~\ref{sec:MICL} details the selection according to the MICL. Section~\ref{sec:missing} focuses on the missing values. 
Section~\ref{sec:experiences} compares the proposed approaches to well-established methods on simulated and illustrates their benefits on real data. Section~\ref{sec:conclude} concludes this work.

\section{Model-based clustering for mixed data} \label{sec:model}
\subsection{The model}
Data to  analyze  consists of $n$ observations $\tx=(\bx_1,\ldots,\bx_n)$, where each observation $\bx_i=(x_{i1},\ldots,x_{id})^\top$ is defined on
space $\mathcal{X}_1\times\ldots\times\mathcal{X}_d$,  
$\mathcal{X}_j$ depending on the nature of variable $j$. Hence,
$\mathcal{X}_j=\mathbb{R}$ (respectively $\mathbb{N}$, $\{1,\ldots,m_j\}$) if
variable $j$ is continuous (respectively integer and categorical with $m_j$
levels). 
Observations are assumed to arise independently from the mixture of $g$ components defined by its probability distribution function (pdf)
\begin{equation}
  \pdf (\bx_i | g, \bth ) = \sum_{k = 1}^g \tau_k \pdf_k(\bx_i |   \balph_k)
   \text{ with }
   \pdf_k(\bx_i |   \balph_k) =\prod_{j = 1}^d \pdf_{kj}(x_{ij} |  \balph_{kj}),
   \label{eq:mixturegrl}
 \end{equation}
 where $\bth=\{\tau_k,\balph_k;k=1,\ldots,g\}$ groups the model parameters, $\tau_k$ is the proportion of
 component $k$ such that $0<\tau_k\leq 1$ and $\sum_{k=1}^g \tau_k=1$, 
 $\pdf_k$ is the pdf of component $k$ parametrized by
 $\balph_k=(\balph_{k1},\ldots,\balph_{kd})$, and  $\pdf_{kj}$ is the pdf of variable $j$ for component $k$
 parametrized by $\balph_{kj}$. The univariate marginal distribution of variable $j$ depends on its definition space, therefore $\pdf_{kj}$ is the pdf of a Gaussian distribution $\mathcal{N}(\mu_{kj},\sigma_{kj}^2)$ (Poisson $\mathcal{P}(\alpha_{kj})$  and multinomial $\mathcal{M}(\alpha_{kj1},\ldots,\alpha_{kjm_j})$) if variable $j$ is continuous (respectively integer and categorical) with $\balph_{kj}=(\mu_{kj},\sigma_{kj})$ (respectively $\balph_{kj}=\alpha_{kj}$ and $\balph_{kj}=(\alpha_{kj1},\ldots,\alpha_{kjm_j})$).

In clustering, a variable is said to be \textit{irrelevant}  if
its univariate margins are invariant over the mixture
components. Considering the model defined by \eqref{eq:mixturegrl}, variable $j$ is irrelevant if $\balph_{1j}=\ldots=\balph_{gj}$, and it is \emph{relevant} otherwise. The role of the variables is defined by the binary vector $\bom=(\omega_1,\ldots,\omega_g)$, since $\omega_j = 0$ if variable $j$ is
irrelevant and $\omega_j = 1$ otherwise. Hence, the couple $\bm=(g,\bom)$ defines the model at hand, because it defines the parameter space. Therefore, for a model $\bm$, the pdf of  $\bx_i$ is
\begin{equation}
\pdf (\bx_i | \bm, \bth ) = \prod_{j \in \bOm^c} \pdf_{1j}(x_{ij} |  \balph_{1j}) \sum_{k = 1}^g \tau_k \prod_{j \in \bOm} \pdf_{kj}(x_{ij} |  \balph_{kj}),
   \label{eq:mixture}
 \end{equation}
 where $\bOm=\{j: \omega_j=1\}$ and $\bOm^c=\{1,\ldots,d\}\setminus \bOm$.
 
\subsection{Maximum likelihood inference}
The general form of the observed-data log-likelihood of  model $\bm$ is defined by $\ell(\bth|\bm,\tx)= \sum_{i=1}^n\ln \left( \sum_{k=1}^g \tau_k\prod_{j=1}^d\pdf_{kj}(x_{ij} |  \balph_{kj})\right)$. Hence, equalities between the parameters defined by $\bom$ imply that
\begin{equation}
\ell(\bth|\bm,\tx)=\left(\sum_{j\in\bOm^c}\sum_{i=1}^n\ln \pdf_{1j}(x_{ij} |  \balph_{1j})\right) + \left( \sum_{i=1}^n\ln \left( \sum_{k=1}^g \tau_k\prod_{j\in\bOm}\pdf_{kj}(x_{ij} |  \balph_{kj})\right)\right).
\end{equation}
The MLE of the parameters corresponding to the irrelevant variables are explicit, but not those of the proportions and the relevant variables. Thus, it is standard to use an EM algorithm \citep{Dem77,McLachlan:08} to maximize the observed-data log-likelihood. Here, the partition among the observations is unobserved. We denote this partition by $\tz=(\bz_1,\ldots,\bz_n)$ with $\bz_i=(z_{i1},\ldots,z_{ig})$, where $z_{ik}=1$ if observation $i$ arises from component $k$ and $z_{ik}=0$ otherwise. Hence, the complete-data likelihood of  model $\bm$ (log-likelihood computed on the observed and unobserved variables) is defined by
\begin{equation}
\ell(\bth|\bm,\tx,\tz)=  \sum_{j\in\bOm^c} \sum_{i=1}^n \ln \pdf_{1j}(x_{ij} |  \balph_{1j})+ \sum_{k=1}^g \sum_{i=1}^n z_{ik} \ln \tau_k +   \sum_{j\in\bOm} \sum_{k=1}^g \sum_{i=1}^n z_{ik} \ln \pdf_{kj}(x_{ij} |  \balph_{kj}).
\end{equation}

The EM algorithm alternates between two steps: the Expectation step (E-step) consisting in computing the expectation of the complete-data likelihood under the current parameters, and the maximization step (M-step) consisting in maximizing this expectation over the model parameters. Thus, this algorithm starts from the initial value of the model parameter $\bth^{[0]}$ randomly sampled and its iteration $[r]$ is defined by\\
\textbf{E-step} Computation of the fuzzy partition $t_{ik}^{[r]}:=\mathbb{E}[Z_{ik}|\bx_i,\bm,\bth^{[r-1]}]$, hence
$$t_{ik}^{[r]}:=\dfrac{\tau_k^{[r-1]} \prod_{j =1}^d \pdf_{kj}(x_{ij} |  \balph_{kj}^{[r-1]})}{\sum_{\ell = 1}^g \tau_\ell^{[r-1]} \prod_{j =1}^d \pdf_{\ell j}(x_{ij} |  \balph_{\ell  j}^{[r-1]})},$$
\textbf{M-step} Maximization of the expected value of the complete-data log-likelihood over the parameters,
$$\tau_k^{[r]}=\dfrac{n_k^{[r]}}{n} \text{ and } \balph_{kj}^{[r]}=\left\{ \begin{array}{rl}
\balph^{\star [r]}_{jk} & \text{if } \omega_j=1 \\
\tilde{\balph}_{1j} & \text{otherwise } \\
\end{array} \right.,$$
where $n_k^{[r]}=\sum_{i=1}^nt_{ik}^{[r]}$, $\tilde{\balph}_{1j}=\argmax_{\balph_{1j}} \sum_{i=1}^n \ln \pdf_{1j}(x_{ij} |  \balph_{1j})$ is the MLE for an irrelevant variable, and 
$\balph^{\star [r]}_{jk}=\argmax_{\balph_{kj}} \sum_{i=1}^n  t_{ik}^{[r]} \ln \pdf_{kj}(x_{ij} |  \balph_{kj})$  is the estimate for an relevant variable.
This algorithm converges to a local optimum of the observed-data log-likelihood. Thus, the MLE for the model $\bm$, denoted by $\hat{\bth}_{\bm}$, is obtained by performing many different random initializations.

\section{Model selection by optimizing the BIC} \label{sec:BIC}
\subsection{Information criterion for data modelling}
 Model selection generally aims to find  the model $\mhat$ which obtains the greatest posterior probability, among a collection  of competing models $\M$.
The number of components of the
competing models is usually bounded by a value $g_{\max}$. Thus,
\begin{equation}
\M=\left\{\bm=(g,\bom ):\; g\in\{1,\ldots,g_{\max}\} \text{ and } \bom \in \{0,1\}^d \right\}.
\end{equation}
By assuming uniformity for the prior distribution of $\bm$, 
$\mhat$ maximizes the integrated likelihood defined by
\begin{equation}
\mhat  = \argmax_{\bm\in\M} p(\tx|\bm)
\text{ with }
p(\tx | \bm) =\int_{\bTh_{\bm}} p(\tx | \bm, \bth) p(\bth | \bm) d\bth,
\end{equation}
where $\bTh_{\bm}$ is the parameter space of model $\bm$,
$p(\tx | \bm, \bth)=\prod_{i=1}^n f(\bx_i | \bm, \bth)$ is the
likelihood function, and $ p(\bth | \bm)$ is the pdf of the prior
distribution of the parameters. Unfortunately, the integrated likelihood is intractable, but many methods permit approximations of its value \citep{Fri12}. The most popular approach consists of using the BIC \citep{Schwarz:78,Ker00}, which approximates the logarithm of the integrated likelihood by Laplace approximation, and thus requires MLE. The BIC is defined by
\begin{equation}
\BIC(\bm)=\ln p(\tx|\bm,\hat{\bth}_{\bm}) - \dfrac{\nu_{\bm}}{2} \ln n,
\end{equation}
where  $\nu_{\bm}$ is the number of independent parameters required by $\bm$.

\subsection{Optimizing the penalized likelihood}
For a fixed number of components $g$, selecting the variables necessitates the comparison of $2^d$ models. Therefore, an exhaustive approach  approximating the integrated likelihood for each competing model is not feasible. Instead, \citet{Raftery:06} carry out  model selection by deterministic algorithms (like a \emph{stepwise} method). However, this approach cannot ensure that the model maximizing the BIC is obtained. Moreover, it can be computationally expensive if many variables are observed. In this paper, model selection is an easier problem, because the model assumes within components independence. This assumption permits the direct maximization of any penalized log-likelihood function defined by
\begin{equation}
\lp(\bth|\bm,\tx) = \ell(\bth|\bm,\tx) - \nu_{\bm} c,
\end{equation}
for any constant $c$. This function is maximized by using a modified version of the EM algorithm \citep{green1990use}. Hence, we introduce the penalized complete-data log-likelihood function
\begin{equation}
\lp(\bth|\bm,\tx,\tz)=  \ell(\bth|\bm,\tx,\tz) - (g-1)c - c\sum_{j=1}^d \nu_j (g \omega_j + 1-\omega_j), 
\end{equation}
where $\nu_j$ is the number of parameters for one univariate marginal distribution of variable $j$ (\emph{i.e.,}, $\nu_j=2$ is the variable is continuous, $\nu_j=1$ is the variable is integer and $\nu_j=m_j-1$ is the variable is categorical with $m_j$ levels). This modified version of the EM algorithm finds the model maximizing the penalized log-likelihood for a fixed number of components. It starts at a initial point $(\bm^{[0]},\bth^{[0]})$ randomly sampled with $\bm^{[0]}=(g,\bom^{[0]})$,  and its iteration $[r]$ is composed of two steps:\\
\textbf{E-step} Computation of the fuzzy partition 
$$t_{ik}^{[r]}:=\dfrac{\tau_k^{[r-1]} \prod_{j =1}^d \pdf_{kj}(x_{ij} |  \balph_{kj}^{[r-1]})}{\sum_{\ell = 1}^g \tau_\ell^{[r-1]} \prod_{j =1}^d \pdf_{\ell j}(x_{ij} |  \balph_{\ell  j}^{[r-1]})},$$
\textbf{M-step} Maximization of the expectation of the penalized complete-data log-likelihood over $(\bom,\bth)$, hence
$\bm^{[r]}=(g,\bom^{[r]})$ with
$$\omega_j^{[r]}=\left\{ \begin{array}{rl}
1 & \text{if } \Delta_j^{[r]} > 0 \\
0 & \text{otherwise}
\end{array}\right.,\; \tau_k^{[r]}=\dfrac{n_k^{[r]}}{n} \text{ and } \balph^{[r]}_{jk}=\left\{ \begin{array}{rl}
\balph^{\star [r]}_{kj} & \text{if } \omega_j^{[r]}=1 \\
\tilde{\balph}_{kj} & \text{otherwise}
\end{array}\right.,$$
where $\Delta_j=\sum_{k=1}^g \sum_{i=1}^n t_{ik}^{[r]} \big(\ln \pdf_{kj}(x_{ij} |  \balph^{\star [r]}_{kj})- \ln\pdf_{1j}(x_{ij} |  \tilde{\balph}_{1j})\big) - (g-1)\nu_j c$ is the difference between the maximum of the expected value of the penalized complete-data log-likelihood obtained when variable $j$ is relevant and when it is irrelevant. To obtain the couple $(\bom,\bth)$ maximizing the penalized observed-data log-likelihood, for a fixed number of components, many random initializations of this algorithm should be done. Hence, the couple $(\bm,\bth)$ maximizing the penalized observed-data log-likelihood is obtained by performing this algorithm for every values of $g$ between one and $g_{\max}$. By considering $c=\dfrac{1}{2}\ln n$, this algorithm carry out the model selection according to the BIC. Moreover, other criteria can also be considered like the AIC by setting $c=1$.

\section{Model selection by optimizing the MICL}\label{sec:MICL}
\subsection{Information criterion}
Although the BIC has good properties of consistency, it does not focus the clustering goal. Moreover, it involves an approximation in $\mathcal{O}(1)$ which can deteriorate its performances, especially when $n$ is small or when $\M$ is large. To circumvent this issue, exact criteria could be preferred \citep{Biernacki:10}. Criteria based on the complete-data likelihood have been introduced like the ICL \citep{Biernacki:00} or the MICL \citep{marbac2015variable}.
The \emph{integrated complete-data likelihood} is defined by
\begin{equation}
p(\tx, \tz |\bm) = \int_{\bTh_{\bm}} p(\tx, \tz |\bm,\bth) p(\bth | \bm) d\bth.
\end{equation}
where $ p(\tx, \tz |\bm,\bth)=\prod_{i=1}^n\prod_{k=1}^g[\tau_kf_k(\bx_i|\balph_k)]^{z_{ik}}$ is the complete-data likelihood. When conjugate prior distributions are used, the integrated
complete-data likelihood has the following closed form. Thus,  we assume independence between the
prior distributions, such that
\begin{equation}
p(\bth | \bm)  = p(\btau | \bm) \prod_{j=1}^d p(\balph_{\bullet j}|g,\omega_j)
\text{ with } p(\balph_{\bullet j}|g,\omega_j)=\left\{
\begin{array}{rl}
\prod_{k=1}^g p(\balph_{kj}) & \text{if } \omega_j=1\\
p(\balph_{1j})\prod_{k=1}^g \mathds{1}_{\{\balph_{kj}=\balph_{1j}\}} & \text{if } \omega_j=0,\\
\end{array}\right.,
\end{equation}
where $\balph_{\bullet j}=(\balph_{1j},\ldots,\balph_{gj})$. We use conjugate prior distributions, thus $\btau|\bm$ follows a
Dirichlet distribution $\mathcal{D}_g(u,\ldots,u)$. If variable $j$ is continuous, $p(\balph_{kj})=p(\sigma_{kj}^2)p(\mu_{kj}|\sigma_{kj}^2)$ where $\sigma_{kj}^2$ follows an Inverse-Gamma distribution $\mathcal{IG}(a_j/2,b_j^2/2)$ and
$\mu_{kj} | \bm, \sigma_{kj}^2$ follows a Gaussian distribution $\mathcal{N}(c_j, \sigma_{kj}^2/d_j)$. If variable $j$ is integer, then $\balph_{kj}$ follows a Gamma distribution $\mathcal{G}a(a_j,b_j)$ while $\balph_{kj}$ follows a Dirichlet distribution $\mathcal{D}_{m_j}(a_{j},\ldots,a_{j})$ if variable $j$ is ordinal with $m_j$ levels. If there is no information \emph{a priori} on the parameters, we use the 
the Jeffreys non informative prior distributions \citep{Rob07} for the proportions (\emph{i.e.,} $u_{k}=1/2$) and for the hyper-parameters of a categorical variables (\emph{i.e.,} $a_{jk}=1/2$). Such prior distributions do not exist for the parameters of the Gaussian and Poisson distributions, so we use flat prior distributions (see Section~\ref{sec:experiences}). 

The conjugate prior distributions implies the following closed-form of the integrated complete-data likelihood
\begin{equation}
p(\tx, \tz | \bm) = 
\frac{\Gamma\left(\frac{g}{2}\right)}{\Gamma\left(\frac{1}{2}\right)^g}\dfrac{\prod_{k=1}^g \Gamma\left(n_k + \frac{1}{2}\right)}{\Gamma\left(n+\frac{g}{2}\right)}
 \prod_{j=1}^d p(\tx_{\bullet j}|g,\omega_j,\tz), \label{eq::prod}
\end{equation}
where $\tx_{\bullet j}=(x_{ij};i=1,\ldots,n)$, $n_k=\sum_{i=1}^nz_{ik}$ and  
\begin{equation}
p(\tx_{\bullet j}|g,\omega_j,\tz)=\int p(\balph_{\bullet j}|g,\omega_j)\prod_{k=1}^g  \prod_{i=1}^nf_{kj}(x_{ij}|\balph_{kj})^{z_{ik}}d\balph_{\bullet j}.\label{eq::integral}
\end{equation}
The integral defined by \eqref{eq::integral} is explicit, thus providing a closed-form of the integrated complete-data likelihood. Its value depends on $\omega_j$ and its form is detailed in Appendix~\ref{sec:integral}.

The MICL corresponds to the greatest value of the integrated complete-data likelihood among all 
the possible partitions. Thus, the MICL  is defined by
\begin{equation}
\MICL(\bm) = \ln p(\tx, \tzstar_{\bm} | \bm) \text{ with } \tzstar_{\bm}=\argmax_{\tz} \ln p(\tx, \tz | \bm).
\end{equation}

Obviously, this criterion is quite similar to the ICL  and inherits its
main properties. In particular, it is less sensitive to model misspecification than the BIC. Unlike the ICL and the BIC, it does not require the MLE and thus avoid the multiple calls to the EM algorithm. Because $\bom$ does not impact the dimension of $\tz$, we can maximize the integrated complete-data likelihood over $(\bom,\tz)$,  and thus the best model according the MICL can be obtained, for a fixed number of components .

\subsection{Optimizing the MICL}
An iterative algorithm is used for finding the model maximizing the MICL, for a fixed number of components. Starting at the initial point $(\tz^{[0]}, \bm^{[0]})$ with $\bm^{[0]}=(g,\bom^{[0]})$, the algorithm alternates between two optimizations of the integrated complete-data likelihood: optimization on $\tz$ given $(\tx,\bm)$, and maximization on $\bom$ given $(\tx,\tz)$.
The algorithm is initialized as follows:  $\omega_j^{[0]}=1$ with probability 0.5 then $\tz^{[0]}=\hat{\tz}_{\bm^{[0]}}$ is the partition provided by a MAP rule associated to model $\bm^{[0]}$ and to its MLE $\hat{\bth}_{\bm^{[0]}}$. Iteration $[r]$ of the algorithm is written as\\
\textbf{Partition step:} find $\tz^{[r]}$ such that 
$$\ln p(\tx, \tz^{[r]} | \bm^{[r]}) \geq \ln p(\tx, \tz^{[r-1]} | \bm^{[r]}) .$$
\textbf{Model step:} find $\bm^{[r+1]}=\argmax_{\bm\in\M_g}
  \ln p(\tx,\tz^{[r]} |\bm)$ such that 
$$\bm^{[r+1]}=(g,\bom^{[r+1]}) \text{ with } \omega_j^{[r+1]} =\left\{ \begin{array}{rl}
1 & \text{if } p(\tx_{\bullet j} |g,\omega_j=1, \tz^{[r]})>p(\tx_{\bullet j} |g,\omega_j=0, \tz^{[r]}) \\
0 & \text{otherwise}
\end{array}\right. .
$$

At iteration $[r]$, the model step consists in finding the vector $\bm^{[r+1]}$ maximizing the integrated completed-data likelihood, for the current partition $\tz^{[r]}$. This optimization can be performed independently for each element $\omega_j$, thanks to the within component independence assumption.
The partition step is more complex, hence  $\tz^{[r]}$ is defined as a partition increasing the value of the  integrated complete-data likelihood for the current model. It is obtained by an iterative method initialized at the partition $\tz^{[r-1]}$. Each iteration consists in sampling uniformly an individual which is affiliated to the 
component maximizing the integrated complete-data likelihood, while the other component memberships are unchanged (details are given in \citet{marbac2015variable}).
Like the EM algorithm, the proposed algorithm converges to a local optimum of $\ln p(\tx, \tz | \bm)$, so many
different initializations should be done.

\section{Missing values}\label{sec:missing}
Data can contain missing values, so we denoted by
 $\mO_i\subseteq\{1,\ldots,d\}$ the indices where $\bx_i$ is observed. Assuming that
 data are missing at random, the pdf of $\bx_i$ is defined by the marginal pdf of the observed values given by
\begin{equation}
\pdf (\bx_i | \bm, \bth ) = \prod_{j \in \bOm^c\cap\mO_i} \pdf_{1j}(x_{ij} |  \balph_{1j}) \sum_{k = 1}^g \tau_k \prod_{j \in \bOm\cap\mO_i} \pdf_{kj}(x_{ij} |  \balph_{kj}).
   \label{eq:mixturemissing}
 \end{equation}

The EM algorithm maximizing the BIC can be used on data with missing values. Its M-step should considers only the observed values. Moreover, the within component independence avoids the computation the conditional expected values of the missing observations, at the E-step. The steps of this algorithm are detailed in Appendix~\ref{sec:EMmissing}. Alternatively, variables can be selected according to the MICL. Note that this criterion is particularly relevant in this case, because it considers the number of missing values in the sample, while the penalty of the BIC neglects this quantity. Indeed, the integrated complete-data considers a number of observations per variable because
\begin{equation}
p(\tx_{\bullet j}|g,\omega_j,\tz)=\int p(\balph_{\bullet j}|g,\omega_j)\prod_{k=1}^g  \prod_{\{i: j\in\mO_i\}}f_{kj}(x_{ij}|\balph_{kj})^{z_{ik}}d\balph_{\bullet j}.\label{eq::integralmissing}
\end{equation}
This integral keeps a closed-form detailed in Appendix~\ref{sec:integralmissing}.

\section{Numerical experiments}\label{sec:experiences}
\paragraph{Implementation of the proposed method} 
%  Results of our method  are provided by the R package \emph{VarSelLCM.2.0} %(Variable Selection for the Latent Class Model) available at \url{https://%r-forge.r-project.org/R/?group_id=2011}, with its options by default. 
Our method is implemented by the name \textit{VarSelLCM}.
When the MICL is used, the hyper-parameters must be specified. For the proportions and the parameters of the categorical data, we use the Jeffrey's prior distributions (\emph{i.e.,} Dirichlet distribution with parameters equal to 1/2).  Because there do not exist non-informative Jeffrey's prior distributions for the Gaussian mixture, the following hyper-parameters are chosen to be fairly  flat in the region where the likelihood is substantial and not much greater else-where:
  $a_j=1$, $b_j=1$,
  $c_j=\text{mean}(\tx_{\bullet j})$ and
  $d_j=0.01$. In the same spirit, we use the hyper-parameters $a_j=b_j=1$ for the Poisson distribution. 
  The purpose of these experiments is to show the relevance of selecting variables in clustering.
Two families of information criteria are considered: the model-fitting criterion (BIC) or the clustering-task criterion.
When we apply the clustering-task criterion, the ICL is used if there is no selection of the variables, while the MICL is used if variables are selected.

\paragraph{Simulation map}
First, methods of variables selection are compared for a model-based clustering of continuous data.
Thus, we compare our approaches with the selection of variables by using  deterministic algorithm maximizing the BIC implemented in the R package clustvarsel \citep{Scr14}. This package considers redundant variables and different covariance matrices. The optimisation of the BIC is proposed by two algorithms: forward and backward searches. Note that there already exist comprehensive comparisons of method for selecting variables in a continuous data clustering \citep{Cel14,marbac2015variable}. Second, the impact of the missing values is illustrates on mixed simulated data. Third, the benefits of the proposed approaches are illustrates on five real datasets.
In this section, method are compared in a clustering task. Thus, the partitioning accuracy is measured with the Adjusted Rand Index (ARI, \citet{Hub85}) because it permits the comparison between two partitions having possibly different numbers of classes. When it is close to one, the partitions are strongly similar, while they are strongly different when this index is close to zero.

\subsection{Simulated data: continuous case}
This experiment compares the methods of model selection on clustering of continuous data.
We generate 200 observations from a bi-component Gaussian mixture with equals proportions. Under component $k$, the $r=6$ variables follow a Gaussian distribution $\mathcal{N}(\bmu_k,\bSigma_k)$ with
$$
\mu_{kj} = \left\{
\begin{array}{cc}
(-1)^k \delta & \text{if } j\leq 6 \\ 
0 & \text{otherwise} \\ 
\end{array} 
\right.
\text{ and }
\bSigma_k[j,j']=\left\{
\begin{array}{cc}
1  & \text{if } j=j' \\ 
\rho  & \text{if } |j-j'|=1 \\
0 & \text{otherwise}\\ 
\end{array} 
\right. .
$$
Where $\delta$ is used to define the class overlap.
We add $d-r$ noisy variables from a standard Gaussian distribution $\mathcal{N}(\boldsymbol{0},\boldsymbol{I})$.
We consider different numbers of variables ($d=$10, 25, 50, 100), a theoretical misclassification rates of 5\% and two values of $\rho$ (0 and 0.4). Thus, when $\rho=0$, the model used for sampling the data belongs to the list of the competing models while it does not belong to this list when $\rho=0.4$ For each case, 20 replicates are generated. 

For each replicates,  we perform the clustering, with unknown number of classes, according to a modelling criterion (BIC) and to a clustering criterion (ICL/MICL). Model selection is performed by considering a maximum number of  components equals to three. The ARI is computed for each selected model and their values are presented in Table~\ref{tab:res1}.

\begin{table}[ht!]
\begin{center}
\begin{tabular}{cc|cccc|cc}
\hline
 &  & \multicolumn{4}{c|}{BIC} & \multicolumn{2}{c}{ICL/MICL}\\
  $\rho$& $d$& no  & \multicolumn{2}{c}{clustvarsel} & VarSelLCM & no  & VarSelLCM\\
&&selection&forward&backward&&selection&\\
\hline
 \multirow{4}{*}{0} &
  10  & 0.78   &  0.53   &  0.71   &  0.78   &  0.78   &  0.78\\
& 25  & 0.31   &  0.34   &  0.71   &  0.77   &  0.04   &  0.77\\
& 50  & 0.00   &  0.13   &  0.04   &  0.80   &  0.00   &  0.80\\
& 100 & 0.00   &  0.10   &  0.00   &  0.77   &  0.00   &  0.77\\
\hline
 \multirow{4}{*}{0.4}
& 10  & 0.78  &   0.52  &   0.69  &   0.72   &  0.78  &   0.78\\
& 25  & 0.78  &   0.44  &   0.63  &   0.77   &  0.78  &   0.78\\
& 50  & 0.50  &   0.30  &   0.03  &   0.79   &  0.08  &   0.80\\
& 100 & 0.00  &   0.18  &   0.00  &   0.73   &  0.00  &   0.77 \\
\hline
\end{tabular} 
\caption{Mean of the ARI obtained by different methods selecting the variables in a continuous data clustering. \label{tab:res1}}
\end{center}
\end{table}

 For both criterion families, selecting the variables increases the clustering accuracy. Even if the data arise from a model with intra-components dependencies ($\rho=0.4$), the proposed approach, which assumes within component independence, stays relevant. Indeed, it obtains a better ARI that the models implemented in clustvarsel. This phenomenon can be explained by different reasons. First, the information about the intra-component dependency vanishes when the number of irrelevant variables increases. Thus, the results of clustervarsel are deteriorated when $d$ increases. Second, the BIC can imply issues when $d$ increases, due to its approximation term in $\mathcal{O}(1)$. Because clustervarsel considers a richer family of models, it can be more sensitive to its issue. Finally, the independence assumption permits to finds the model maximizing the BIC, while a richer family of models implies a sub-optimal optimization of this criterion.

Table~\ref{tab:res1nis} gives information about the model selected by the competing methods (number of components and rate of relevant variables). It shows that, when $n$ is fixed, the number of components tends to one, when we handle a very large number of irrelevant variables. This problem is circumvented, when the proposed procedure of variable selection is used with the BIC or the MICL. Moreover, this approach permits the detection of the role of the variables. Indeed for $d=10$, 25, 50 and 100, the rate of relevant variables (\emph{i.e.,} 0.60, 0.25, 0.13, 0.06) is found.

\begin{table}[ht!]
\begin{center}
\begin{tabular}{cc|ccccccc|ccc}
\hline
 &  & \multicolumn{7}{c|}{BIC} & \multicolumn{3}{c}{ICL/MICL}\\
  $\rho$& $d$& no & \multicolumn{4}{c}{clustvarsel} & \multicolumn{2}{c|}{VarSelLCM} &no & \multicolumn{2}{c}{VarSelLCM}\\
&&selection&\multicolumn{2}{c}{forward}&\multicolumn{2}{c}{backward}&&&selection&\\
&& $g$& $g$ & rel.& $g$ & rel.& $g$ & rel.& $g$ &  $g$ & rel.\\
\hline
 \multirow{4}{*}{0} 
& 10  &  2.00 & 2.45 & 0.52 & 2.55 & 0.66 &  2.00 & 0.60 & 2.00 &    2.00 &  0.60\\
& 25  &  1.40 & 2.80 & 0.22 & 2.40 & 0.50 &  2.00 & 0.25 & 1.05 &    2.00 &  0.24\\
& 50  &  1.00 & 2.90 & 0.08 & 2.95 & 0.43 &  2.00 & 0.13 & 1.00 &    2.00 &  0.12\\
& 100 &  1.00 & 2.95 & 0.04 & 3.00 & 0.72 &  2.00 & 0.06 & 1.00 &    2.00 &  0.06\\
 \hline
 \multirow{4}{*}{0.4} 
& 10  & 2.00 & 2.60 & 0.28 & 2.55 & 0.40 & 2.25 & 0.61 &  2.00 &    2.00 &  0.60\\
& 25  & 2.00 & 2.85 & 0.17 & 2.65 & 0.40 & 2.05 & 0.25 &  2.00 &    2.00 &  0.24\\
& 50  & 1.65 & 2.85 & 0.12 & 2.85 & 0.44 & 2.05 & 0.13 &  1.10 &    2.00 &  0.12\\
& 100 & 1.00 & 2.95 & 0.04 & 3.00 & 0.70 & 2.15 & 0.06 &  1.00 &    2.00 &  0.06\\
 \hline
\end{tabular} 
\caption{Mean of  component number ($g$) and rate of releveant variables (rel.)  obtained  different methods selecting the variables in a continuous data clustering. \label{tab:res1nis}}
\end{center}
\end{table}

\subsection{Simulated data: mixed case}
This experiment evaluates the benefits of variable selection, for clustering mixed data with missing values, if either a modelling criterion (BIC) or a clustering criterion (ICL/MICL) is used. We generate 200 observations  from a bi-component model with equals proportions and assuming within components independence. We consider $r=6$ variables (two continuous, two integer and two binary).
Under component $k$, the univariate margins are defined by these parameters
$$
\left\{
\begin{array}{rll}
\mu_{kj}&=(-\delta)^k,\; \sigma_{kj}=1 & \text{for the continuous variables,}\\
\alpha_{kj}&=3+(-\delta)^k & \text{for the integer variables,}\\
\alpha_{kj}&=0.5+(-0.2)^k& \text{for the binary variables.}
\end{array} 
\right. 
$$
The parameter $\delta$ allows us to fix the misclassification error at 10\%. Noisy variables are added (equal number of continuous, integer and binary variables), then  missing values are added randomly. Thus, 20 replicates are generated for different numbers of variables ($d=$12, 24, 48) and different rates of missing values (0\%, 10\% and 20\%). Table~\ref{tab:res2} presents the results.

\begin{table}[ht]
\begin{center}
\begin{tabular}{cc|ccccc|ccccc}
\hline
 &  & \multicolumn{5}{c|}{BIC} & \multicolumn{5}{c}{ICL/MICL}\\
missing& $d$& \multicolumn{2}{c}{no selection} & \multicolumn{3}{c|}{selection} & \multicolumn{2}{c}{no selection} & \multicolumn{3}{c}{selection}\\
values& & ARI & g & ARI & g & rel. & ARI & g & ARI & g & rel.\\
\hline
& 12 & 0.42  &  1.90  &  0.57  &  2.00  &  0.48  &  0.25  &  1.40  &  0.34  &  1.60  &  0.78\\
0\%& 24 & 0.52  &  1.90  &  0.61  &  2.00  &  0.26  &  0.46  &  1.80  &  0.50  &  2.05  &  0.41\\
& 48 & 0.35  &  1.60  &  0.59  &  2.00  &  0.14  &  0.33  &  1.95  &  0.39  &  2.00  &  0.22\\
\hline
& 12 & 0.29  &  2.00  &  0.51  &  2.00  &  0.47  &  0.16  &  1.30  &  0.19  &  1.40  &  0.82\\
10\% & 24 & 0.43  &  2.20  &  0.55  &  2.00  &  0.26  &  0.32  &  1.55  &  0.38  &  1.80  &  0.60\\
& 48 & 0.16  &  2.00  &  0.52  &  2.00  &  0.13  &  0.13  &  1.80  &  0.17  &  2.05  &  0.38\\
\hline
& 12 & 0.12  &  2.10  &  0.43  &  2.00  &  0.44  &  0.05  &  1.10  &  0.05  &  1.10  &  0.94\\
20\% & 24 & 0.20  &  2.30  &  0.48  &  2.00  &  0.24  &  0.13  &  1.30  &  0.15  &  1.40  &  0.79\\
& 48 & 0.08  &  2.00  &  0.41  &  2.00  &  0.12  &  0.05  &  1.85  &  0.09  &  2.05  &  0.35\\
\hline
\end{tabular} 
\caption{Results of the different approaches to cluster mixed data considering differents numbers of variables and rates of missing values (misclassification rate of 10\%): mean of the ARI (ARI), mean of the number of components (g) and mean of the number of relevant variables (rel.).\label{tab:res2}}
\end{center}
\end{table}

 Results shows that the selecting the variables increases the clustering performances for both types of criteria. Indeed, the values of the ARI are improved when variables are selected, especially when the clustering criteria are used. Moreover, the true number of components ($g=2$) is more often found when variables are selected. Finally, clustering interpretation is facilitates because only a subset of the observed variables characterizes the classes. As expected, when missing values are added, the results are deteriorated. However, the results are more impacted when all the variables are used to cluster. Note the BIC obtains better results for detecting the role of the variables. Indeed, the rate of discriminative variables is 0.50, 0.25 and 0.125 when $d$ is equal to 12, 24 and 48 respectively. For this simulation, the overall behaviour of the BIC criterion is better than the one of the MICL. This phenomenon is explained by a quite large class overlaps. It is known that the information criteria based on the integrated complete-data likelihood work better when the classes are well-separated. To illustrate this phenomenon, we perform a similar simulation by considering the 5\% of theoretical misclassification and 20\% of missing values. Table~\ref{tab:res3} shows the results obtained when the variables are selecting according to the BIC and the MICL. In this case, both criteria obtain equivalent results for the ARI, the number of component and the detection of the variables. In this section, the model used for sampling the data belongs to the list of the competing models. This favour the BIC criterion. Next section shows that the MICL is at least as relevant as the BIC for selecting the variables when real data are analysed.
 
\begin{table}[ht]
\begin{center}
\begin{tabular}{c|ccc|ccc}
\hline
& \multicolumn{3}{c|}{BIC}  & \multicolumn{3}{c}{MICL}\\
$d$  & ARI & g & rel.  & ARI & g & rel.\\
\hline
12 &  0.69  &  2.00  &  0.50  &  0.68  &  2.00  &  0.49 \\
24  &  0.69  &  2.00  &  0.27  &  0.65  &  2.00  &  0.30 \\
48  &  0.70  &  2.00  &  0.13  &  0.62  &  2.00  &  0.16  \\
\hline
\end{tabular} 
\caption{Criterion comparison for selecting the variables in a mixed data clustering (misclassification rate of 5\% and 20\% of missing values): mean of the ARI (ARI), mean of the number of components (g) and mean of the number of relevant variables (rel.).\label{tab:res3}}
\end{center}
\end{table}

\subsection{Method comparison on  real data}
  We now compare the competing methods on six real datasets presented in Table~\ref{tab::datasets}.

\begin{table}[ht!]
\begin{center}
\begin{tabular}{cccccc}
\hline 
Name &  $n$ & $d$ & Classes & Reference & R package/website\\ 
\hline 
Birds &  69 & 5 & 2 & \citet{Bretagnolle07} & Rmixmod \\
Banknote &  200 & 6 & 2 & \citet{Flu88} & VarSelLCM \\
 Coffee &  43 & 12 & 2 & \citet{Str73} &  ppgm \\ 
 Statlog-Heart&  43 & 12 & 2 & \citet{Brown04} &   UCI-database  \\ 
 Congress &  435 & 16 & 2 & \citet{Schlimmer:87} &  UCI-database \\ 
 Golub &  83 & 3051 & 2 & \citet{Gol99} & multtest \\ 
\hline \end{tabular} 
\end{center}
\caption{Information about the benchmark datasets. \label{tab::datasets}}
\end{table}

Table~\ref{tab:resGknown} presents the results obtained without variable selection (No selection), with a variable selection according to the BIC (BIC-selection) and with a variable selection according to the MICL (MICL-selection). For three datasets (birds, coffe, banknote), the ARI obtained by the three approaches are equal. However, selecting the variables facilitates the clustering interpretation. For example, the selection with the MICL perfectly detects the clusters with only 42\% of the variables. For the three other datasets, selecting the variables increases the ARI.

\begin{table}[ht!]
\begin{center}
\begin{tabular}{ccccccc}
\hline
Dataset & \multicolumn{2}{c}{No selection} & \multicolumn{2}{c}{BIC-selection} & \multicolumn{2}{c}{MICL-selection} \\
& ARI & rel. & ARI & rel. & ARI & rel. \\
\hline 
Birds & 0.50  &  1.00  & 0.50  &  0.60  &  0.50  &  0.60 \\
Banknote & 0.96  &  1.00  & 0.96  &  0.83  &  0.96  &  0.83 \\
Coffee & 1.00  &  1.00  &  1.00  &  0.67  &  1.00  &  0.42 \\
Statlog-Heart & 0.25  &  1.00  &  0.31  &  0.69  & 0.33  &  0.69 \\
Congress & 0.56  &  1.00  &  0.57 &  0.88  &  0.57  &  0.88 \\
Golub & 0.53  &  1.00  &  0.70  &  0.38  & 0.79  &  0.18 \\
\hline
\end{tabular} 
\caption{Results obtained on the real datasets with known number of components: ARI and proportion of relevant variables (rel.). \label{tab:resGknown}}
\end{center}
\end{table}

In many applications the number of classes is unknown. Thus, we perform the clustering, with unknown number of classes, according to a modelling criterion (BIC) and to a clustering criterion (ICL/MICL). For both approaches, the clustering is done with and without selection of the variables. Table~\ref{tab:resGunknown} presents the results obtained for the real datasets. For both approaches, the selection of the variables increases the ARI for almost all the datasets. The only case when the ARI is deteriorated by a selection of variables is for the clustering of the Golub dataset with the BIC. This dataset is really challenging because it is composed of many variables (3051) and few observations (83). This large number of variables implies a huge number of competing models. In this case, the BIC can lead to poor results because of its approximation implying a term of $\mathcal{O}(1)$. For this type of dataset, the exact criteria (implying no approximation) are more relevant. Thus, the clustering with variable selection according to the MICL criterion finds the true number of components (2) and a relevant partition.

\begin{table}[ht!]
\begin{center}
\begin{tabular}{c|ccccc|ccccc}
\hline
 & \multicolumn{5}{c|}{BIC} & \multicolumn{5}{c}{ICL/MICL} \\
Dataset & \multicolumn{2}{c}{No selection} & \multicolumn{3}{c|}{Selection} & \multicolumn{2}{c}{No selection}& \multicolumn{3}{c}{Selection} \\
& ARI & g &  ARI & g & rel. & ARI & g &  ARI & g & rel.  \\
\hline 
Birds & 0.50  &  2  &    0.50  &  2  &  0.60  &  0.50  &  2  &   0.50  &  2  &  0.60 \\
Banknote & 0.48  &  4  &    0.48  &  4  &  1.00  &  0.61 &  3  &    0.61  &  3  &  1.00 \\
Coffee & 0.38  &  3  &    0.38  &  3  &  0.67  & 1.00 &  2  &    1.00  &  2  &  0.42 \\
Statlog-Heart & 0.25  &  2  &   0.31  &  2  &  0.69  &  0.25  &  2  &    0.33  &  2  &  0.69 \\
Congress & 0.40  &  4  &    0.46  &  4  &  0.88  &  0.47  &  5  &    0.47  &  5  &  0.94 \\
Golub & 0.53  &  2  &    0.32  &  4  &  0.32  &  0.00  &  1  &    0.79  &  2  &  0.18 \\
\hline
\end{tabular} 
\caption{Results of the method comparison with unknown number of components: ARI, best number of components (g) and proportion of relevant variables (rel.). \label{tab:resGunknown}}
\end{center}
\end{table}

\section{Discussion} \label{sec:conclude}
We have proposed a new model-based approach for selecting variables in a cluster analysis of mixed data with missing values. The purpose of the variable selection is to increase the accuracy of the model fitting and facilitate its interpretation. The model at hand assumes within component independence. This assumption is relevant, because variable selection is mainly impacting when several variables are observed. Moreover, numerical experiments have shown robustness properties for the model misspecification. The within component independence assumption permits the maximization of the BIC and the MICL. The first criterion performs the selection of the variables and the clustering in a model-fitting purpose. The second criterion achieves these objectives in a clustering purpose. Both criteria have provided relevant results on numerical experiments. Considering a dataset composed of several variables but very few observations, the MICL should be prefered to the BIC, because it does not imply any approximation. However, if many observations are available, the maximization of the MICL could be time consuming (in practice, it is doable for $n<10^4$). Thus, the BIC could be prefered, in this case, because its issues due to the term $\mathcal{O}(1)$ vanishes when $n$ tends toward infinity.

Finally, this approach could be extend to perform a more elaborate variable selection. Indeed, by using the approach of \citet{Mau09}, a group of redundant variables could be considered. 
%The R package \emph{VarSelLCM}.2.0 implementing the proposed method is available at  \url{https://r-forge.r-project.org/R/?group_id=2011}.

\bibliographystyle{apalike}
\bibliography{biblio}
\appendix
\section{Details on the closed-form of the integrated complete-data log-likelihood} \label{sec:integral}
To compute the  integrated complete-data log-likelihood, we give the value $p(\tx_{\bullet j}|g,\omega_j,\tz)$ for any type of data (continuous, integer and categorical).
\begin{itemize}
\item If variable $j$ is continuous, then
\begin{equation*}
p(\tx_{\bullet j}|g,\omega_j,\tz)=\left\{ \begin{array}{rl}
\pi^{-n/2}
\left( \dfrac{b_j^{a_j/2}d_j^{1/2}}{\Gamma(a_j/2)}\right)^g
\prod_{k=1}^g

\dfrac{\Gamma(A_{kj}/2)}{B_{kj}^{A_{kj}}D_{kj}^{1/2}}

& \text{if } \omega_j = 1 \\
 \pi ^{-n/2} \dfrac{b_j^{a_j}d_j^{1/2}}{\Gamma(a_j/2)}\dfrac{\Gamma(A_j/2)}{B_j^{A_j}D_j^{1/2}}

& \text{if } \omega_j = 0
\end{array}\right.  ,
\end{equation*}
where $A_j=n+a_j$, $B_j^2=b_j^2 + \sum_{i=1}^n (x_{ij} - \bar{\text{x}}_j)^2 + \dfrac{(c_j - \bar{\text{x}}_j)^2}{d_j^{-1} + n^{-1}}$, $D_j=n + d_j$, $\bar{\text{x}}_j=\dfrac{1}{n}\sum_{i=1}^n x_{ij}$,  $A_{kj}=n_k+a_j$, 
$B_{kj}^2=b_j^2 + \sum_{i=1}^n z_{ik} (x_{ij} - \bar{\text{x}}_{jk})^2 + \dfrac{(c_j - \bar{\text{x}}_{jk})^2}{d_j^{-1} + n_k^{-1}}$, $D_{kj}=n_k + d_j$, $\bar{\text{x}}_{jk}=\dfrac{1}{n_k}\sum_{i=1}^n z_{ik}x_{ij}$ and $n_k=\sum_{i=1}^nz_{ik}$.
\item If variable $j$ is integer, then
\begin{equation*}
p(\tx_{\bullet j}|g,\omega_j,\tz)=\left\{ \begin{array}{rl}
\dfrac{1}{\prod_{i=1}^n \Gamma(x_{ij}+1)} \left( \dfrac{b_j^{a_j}}{\Gamma(a_j)}\right)^g \prod_{k=1}^g \Gamma(A_{kj})B_{kj}^{-A_{kj}}

& \text{if } \omega_j = 1 \\
\dfrac{1}{\prod_{i=1}^n \Gamma(x_{ij}+1)} \dfrac{b_j^{a_j}}{\Gamma(a_j)} \Gamma(A_{j})B_{j}^{-A_{j}}

& \text{if } \omega_j = 0
\end{array}\right.  ,
\end{equation*}
where $A_j=\sum_{i=1}^n x_{ij}+a_j$, $B_j=b_j^2 + n$, $A_{kj}=\sum_{i=1}^n z_{ik}x_{ij}+a_j$ and $B_j=b_j^2 + \sum_{i=1}^nz_{ik}$.
\item If variable $j$ is categorical with $m_j$ levels, then
\begin{equation*}
p(\tx_{\bullet j}|g,\omega_j,\tz)=\left\{ \begin{array}{rl}
\left( \dfrac{\Gamma\big(m_j a\big)}{\Gamma(a)^{m_j}} \right)^g
\prod \limits_{k = 1}^g  
     \dfrac{\prod_{h = 1}^{m_j}\Gamma\big(\sum_{i=1}^n z_{ik}\mathds{1}_{\{x_{ij}=h\}}  + a_j\big)}{\Gamma\big(\sum_{i=1}^n z_{ik} + m_j a_j\big)} &    \text{if } \omega_j = 1 \\
 \dfrac{\Gamma\big(m_j a\big)}{\Gamma(a)^{m_j}}

     \dfrac{\prod_{h = 1}^{m_j}\Gamma\big(\sum_{i=1}^n \mathds{1}_{\{x_{ij}=h\}} + a_j\big)}{\Gamma\big(n + m_j a_j\big)}& \text{ if } \omega_j = 0
\end{array}\right.  .
\end{equation*}
\end{itemize}

\section{EM algorithm to optimize the BIC criterion for data with missing values}
\label{sec:EMmissing}
The EM algorithm starts at a initial point $(\bm^{[0]},\bth^{[0]})$ with $\bm^{[0]}=(g,\bom^{[0]})$ randomly sampled and its iteration $[r]$ is composed of two steps:\\
\textbf{E step} Computation of the fuzzy partition 
$$t_{ik}^{[r]}:=\dfrac{\tau_k^{[r-1]} \prod_{j \in \mO_i} \pdf_{kj}(x_{ij} |  \balph_{kj}^{[r-1]})}{\sum_{\ell = 1}^g \tau_\ell^{[r-1]} \prod_{j \in \mO_i} \pdf_{\ell j}(x_{ij} |  \balph_{\ell  j}^{[r-1]})},$$
\textbf{M step} Maximization of the expectation of the penalized complete-data log-likelihood over $(\bom,\bth)$, hence
$\bm^{[r]}=(g,\bom^{[r]})$ with
$$\omega_j^{[r]}=\left\{ \begin{array}{rl}
1 & \text{if } \Delta_j^{[r]} > 0 \\
0 & \text{otherwise}
\end{array}\right.,\; \tau_k^{[r]}=\dfrac{n_k^{[r]}}{n} \text{ and } \balph^{[r]}_{jk}=\left\{ \begin{array}{rl}
\balph^{\star [r]}_{kj} & \text{if } \omega_j^{[r]}=1 \\
\tilde{\balph}_{kj} & \text{otherwise}
\end{array}\right.,$$
where $\Delta_j=\sum_{k=1}^g \sum_{\{i:\;j\in \mO_i \}} t_{ik}^{[r]} \big(\ln \pdf_{kj}(x_{ij} |  \balph^{\star [r]}_{kj})- \ln\pdf_{1j}(x_{ij} |  \tilde{\balph}_{1j})\big) - (g-1)\nu_j c$, where $\tilde{\balph}_{1j}=\argmax_{\balph_{1j}} \sum_{\{i:\;j\in \mO_i \}} \ln \pdf_{1j}(x_{ij} |  \balph_{1j})$ and where $\balph^{\star [r]}_{kj}=\argmax_{\balph_{1j}} \sum_{\{i:\;j\in \mO_i \}} t_{ik}^{[r]}  \ln \pdf_{1j}(x_{ij} |  \balph_{1j})$.

\section{Details on the closed-form of the integrated complete-data log-likelihood for data with missing values} \label{sec:integralmissing}
To compute the  integrated complete-data log-likelihood, for data with missing values we give the value $p(\tx_{\bullet j}|g,\omega_j,\tz)$ for any type of data (continuous, integer and categorical) containing missing values.
\begin{itemize}
\item If variable $j$ is continuous, then
\begin{equation*}
p(\tx_{\bullet j}|g,\omega_j,\tz)=\left\{ \begin{array}{rl}
\pi^{-n_j/2}
\left( \dfrac{b_j^{a_j/2}d_j^{1/2}}{\Gamma(a_j/2)}\right)^g
\prod_{k=1}^g

\dfrac{\Gamma(A_{kj}/2)}{B_{kj}^{A_{kj}}D_{kj}^{1/2}}

& \text{if } \omega_j = 1 \\
 \pi ^{-n_j/2} \dfrac{b_j^{a_j}d_j^{1/2}}{\Gamma(a_j/2)}\dfrac{\Gamma(A_j/2)}{B_j^{A_j}D_j^{1/2}}

& \text{if } \omega_j = 0
\end{array}\right.  ,
\end{equation*}
where $n_j=\sum_{i=1}^n \mathds{1}_{\{j\in \mO_i \}}$, $A_j=n_j+a_j$, $B_j^2=b_j^2 + \sum_{\{i:\;j\in \mO_i \}} (x_{ij} - \bar{\text{x}}_j)^2 + \dfrac{(c_j - \bar{\text{x}}_j)^2}{d_j^{-1} + n_j^{-1}}$, $D_j=n_j + d_j$, $\bar{\text{x}}_j=\dfrac{1}{n_j}\sum_{\{i:\;j\in \mO_i \}} x_{ij}$,  $A_{kj}=n_{jk}+a_j$, 
$B_{kj}^2=b_j^2 + \sum_{\{i:\;j\in \mO_i \}} z_{ik} (x_{ij} - \bar{\text{x}}_{jk})^2 + \dfrac{(c_j - \bar{\text{x}}_{jk})^2}{d_j^{-1} + n_{jk}^{-1}}$, $D_{kj}=n_{jk} + d_j$, $\bar{\text{x}}_{jk}=\dfrac{1}{n_{jk}}\sum_{\{i:\;j\in \mO_i \}} z_{ik}x_{ij}$ and $n_{jk}=\sum_{\{i:\;j\in \mO_i \}}z_{ik}$.
\item If variable $j$ is integer, then
\begin{equation*}
p(\tx_{\bullet j}|g,\omega_j,\tz)=\left\{ \begin{array}{rl}
\dfrac{1}{\prod_{\{i:\;j\in \mO_i \}} \Gamma(x_{ij}+1)} \left( \dfrac{b_j^{a_j}}{\Gamma(a_j)}\right)^g \prod_{k=1}^g \Gamma(A_{kj})B_{kj}^{-A_{kj}}

& \text{if } \omega_j = 1 \\
\dfrac{1}{\prod_{\{i:\;j\in \mO_i \}} \Gamma(x_{ij}+1)} \dfrac{b_j^{a_j}}{\Gamma(a_j)} \Gamma(A_{j})B_{j}^{-A_{j}}

& \text{if } \omega_j = 0
\end{array}\right.  ,
\end{equation*}
where $A_j=\sum_{\{i:\;j\in \mO_i \}}+a_j$, $B_j=b_j^2 + n_j$, $A_{kj}=\sum_{\{i:\;j\in \mO_i \}} z_{ik}x_{ij}+a_j$, $B_j=b_j^2 + \sum_{\{i:\;j\in \mO_i \}}z_{ik}$ and $n_j=\sum_{i=1}^n \mathds{1}_{\{j\in \mO_i \}}$.
\item If variable $j$ is categorical with $m_j$ levels, then
\begin{equation*}
p(\tx_{\bullet j}|g,\omega_j,\tz)=\left\{ \begin{array}{rl}
\left( \dfrac{\Gamma\big(m_j a\big)}{\Gamma(a)^{m_j}} \right)^g
\prod \limits_{k = 1}^g  
     \dfrac{\prod_{h = 1}^{m_j}\Gamma\big(\sum_{\{i:\;j\in \mO_i \}} z_{ik}\mathds{1}_{\{x_{ij}=h\}}  + a_j\big)}{\Gamma\big(\sum_{\{i:\;j\in \mO_i \}} z_{ik} + m_j a_j\big)} &    \text{if } \omega_j = 1 \\
 \dfrac{\Gamma\big(m_j a\big)}{\Gamma(a)^{m_j}}

     \dfrac{\prod_{h = 1}^{m_j}\Gamma\big(\sum_{\{i:\;j\in \mO_i \}} \mathds{1}_{\{x_{ij}=h\}} + a_j\big)}{\Gamma\big(\sum_{i=1}^n \mathds{1}_{\{j\in \mO_i \}} + m_j a_j\big)}& \text{ if } \omega_j = 0
\end{array}\right. .
\end{equation*}
\end{itemize}

\end{document}